\documentclass{iau}
\usepackage{graphicx}

\title[IAUS~304.~~Supernovae in paired galaxies] 
{Supernovae in paired galaxies}

\author[T.~A.~Nazaryan et al.]   
{T.~A.~Nazaryan$^1$,
A.~R.~Petrosian$^1$,
A.~A.~Hakobyan$^1$,\\
V.~Zh.~Adibekyan$^2$,
D.~Kunth$^3$,
G.~A.~Mamon$^3$,
M.~Turatto$^4$,\\
 \and L.~S.~Aramyan$^1$}

\affiliation{$^1$Byurakan Astrophysical Observatory, 0213 Byurakan, Aragatsotn Province, Armenia \\
email: {\tt nazaryan@bao.sci.am} \\[\affilskip]
$^2$Centro de Astrof\'{i}sica da Universidade do Porto, Rua das Estrelas, 4150-762 Porto, Portugal\\[\affilskip]
$^3$Institut d'Astrophysique de Paris, 98 bis Bd Arago, 75014 Paris, France\\[\affilskip]
$^4$INAF-Osservatorio Astronomico di Padova, Vicolo dell'Osservatorio 5, 35122 Padova, Italy}

\pubyear{2013}
\volume{304}  
\pagerange{000--000}
\jname{Multiwavelength AGN Surveys and Studies}
\editors{A.~M.~Mickaelian \& D.~B.~Sanders, eds.}
\begin{document}

\maketitle

\begin{abstract}
We investigate the influence of close neighbor galaxies on the properties of supernovae (SNe) and their host galaxies using 56 SNe located in pairs of galaxies with different levels of star formation (SF) and nuclear activity. The mean distance of type II SNe from nuclei of hosts is greater by about a factor of 2 than that of type Ibc SNe. The distributions and mean distances of SNe are consistent with previous results compiled with the larger sample. For the first time it is shown that SNe Ibc are located in pairs with significantly smaller difference of radial velocities between components than pairs containing SNe Ia and II. We consider this as a result of higher star formation rate (SFR) of these closer systems of galaxies.
\end{abstract}


Interaction with a neighbor galaxy can be a triggering mechanism for
nuclear activity and/or circumnuclear starburst
(\cite[Barton et al. 2000]{barton00},
\cite[Ellison et al. 2008]{ellison08}).
Observational results (e.g.
\cite[Cappellaro et al. 1999]{cappellaro99},
\cite[Mannucci et al. 2005]{mannucci05},
\cite[Petrosian et al. 2005]{petrosian05},
\cite[Habergham et al. 2012]{habergham12})
suggest that core-collapse (CC) SNe are tightly connected with recent SF.
The aim of this study is to investigate to what extent
gravitational interaction with a close neighbor can be
connected with nuclear activity and/or enhanced SF in galaxy pairs,
using SNe as tracers of recent SF.
The complete study is presented in \cite{nazaryan13}.

The sample of the current study was obtained by cross-matching
a sample of selected pairs of galaxies with the catalog of SNe by \cite{hakobyan12}
containing 3876 SNe located within the SDSS DR8 coverage.
We used three catalogs of galaxies with different levels of nuclear activity
to construct our sample of close pairs of galaxies.
These catalogs are the following:
(1) the catalog of Markarian (MRK) galaxies,
(2) the Second Byurakan Survey (SBS) galaxies catalog, and
(3) the North Galactic Pole (NGP) galaxy catalog.
Results of a close neighbors search for MRK galaxies
are published (\cite[Nazaryan et al. 2012]{nazaryan12}).
We also conducted the search of neighbors for SBS and NGP galaxies using the same criteria.
For the current study, we identified 56 SNe in 44 hosts
(19 SNe Ia, 12 SNe Ibc, 15 SNe II, and 10 unclassified).
Pairs of galaxies are presented via two parameters
describing strength/stage of interacting/merging:
difference of radial velocities ${\rm d}V_r$
and linear projected distance $D\rm{p}$ between pair members.

Hosts of type Ibc and II SNe in our sample (with median \emph{t}-types Sbc and Sc respectively)
are of later morphological classes than those of type Ia.
This is well known observational result (\cite[Cappellaro et al. 1999]{cappellaro99}).
Morphologies of hosts from our sample are consistent with those of the unbiased
sample of 1021 nearby hosts of different types SNe from \cite{hakobyan12}.
SN hosts of our sample tend to form pairs with neighbors with similar morphologies.
The percentage of barred galaxies among our hosts is larger ($2 \sigma$)
than that in the sample from \cite{hakobyan12}.
Since our sample consists of paired hosts only, the excess of bars is expected
(e.g. \cite[M{\'e}ndez-Abreu et al. 2012]{mendezabreu12}).

We studied radial distributions of SNe of different types.
The mean value of $R_{\rm SN}/R_{25}$  of is $0.53 \pm 0.10$ for type Ia.
The mean normalized distance $R_{\rm SN}/R_{25}$ of SNe II,
is roughly double that of SNe Ibc (with $3.35 \sigma$ level of significance).
Our mean values for CC SNe within estimated errors are in agreement with those of \cite{hakobyan08} and \cite{hakobyan09}.
The higher type II to Ibc ratio of mean normalized distances in our sample
is marginally significant ($1.54 \sigma$) than that in the \cite{hakobyan09} sample.

We found that
the distributions of ${\rm d}V_r$  of pairs containing SNe Ia and II
is the same with practically consistent mean values (about $130~km~s^{-1}$).
But the same distributions of pairs with Ibc and II SNe are significantly different ($ 3.2 \sigma$)
with smaller mean ${\rm d}V_r$  of for Ibc ($56~km~s^{-1}$ in average).
However, there is no difference between mean values of $D\rm{p}$
of pairs with SNe of different types.
To explain the strong dependence of CC SNe types on ${\rm d}V_r$,
we considered SFR as the main parameter affecting SN production in galaxies.
SN rate and number ratio of type Ibc to type II SNe should increases with increasing of SFR
(\cite[Cappellaro et al. 1999]{cappellaro99},
\cite[Mannucci et al. 2005]{mannucci05},
\cite[Petrosian et al. 2005]{petrosian05},
\cite[Hakobyan et al. 2011]{hakobyan11}).
Also we expect a relatively larger amount of star-forming galaxies
in pairs with smaller ${\rm d}V_r$ and $D\rm{p}$
due to interaction-triggered starbursts
(\cite[Barton et al. 2000]{barton00},
\cite[Ellison et al. 2008]{ellison08}).
Therefore, the excess of Ibc SNe compared to II SNe in the pairs
with smaller ${\rm d}V_r$ and $D\rm{p}$ can be a result of a higher SFR in their stronger interacting hosts.

As a conclusion, we consider that close environment of galaxies can have
some observable effect on SN production due to the impact on SF of galaxies.

\begin{acknowledgement}
A.~R.~P., A.~A.~H., and L.~S.~A. are supported by the Collaborative Bilateral Research Project
of the State Committee of Science (SCS) of the Republic of Armenia
and the French Centre National de la Recherch\'e Scientifique (CNRS).
This work was made possible in part by a research grant from the
Armenian National Science and Education Fund (ANSEF) based in New York, USA.
This work was supported by State Committee Science MES RA,
in frame of the research project number SCS 13-1C013.
V.Zh.A. is supported by grant SFRH/BPD/70574/2010 from FCT (Portugal)
and would further like to thank for the support by the ERC under the FP7/EC
through a Starting Grant agreement number 239953.
\end{acknowledgement}

\end{document}